\documentclass[a4paper,11pt]{article}
\pdfoutput=1 

\usepackage{jinstpub} 

\usepackage[bottom,flushmargin,symbol]{footmisc}

\title{Surface background rejection technique for liquid argon dark matter detectors using a thin scintillating layer}

\author{D. Gallacher\footnote{Corresponding Author}}
\author{and M. Boulay}
\affiliation{Carleton University,\\1125 Colonel By Dr, Ottawa ON, Canada}

\emailAdd{david.gallacher@carleton.ca}

\abstract{Future large liquid argon direct dark matter detectors can benefit greatly from an efficient surface background rejection technique. To aid the development of these large scale detectors a test stand, Argon-1, has been constructed at Carleton University, Ottawa, Canada, in the noble liquid detector development lab. It aims to test a novel surface background rejection technique using a thin layer of slow scintillating material at the surface of the vessel. Through pulse-shape discrimination of the slow light from the scintillating layer, events from the surface of the detector can be discriminated from liquid argon events. The detector will be implemented with high-granularity SiPMs for light detection which will be used to accurately identify surface events to characterize the proposed technique. An overview of the technique and  the status of the experiment are discussed here. }


\arxivnumber{1911.07265} 

\proceeding{LIDINE 2019: Light Detection In Noble Elements\\
  28-30 August 2019\\
  University of Manchester, UK}

\begin{document}
\maketitle
\flushbottom

\section{DEAP-3600 and Beyond}
\label{sec:intro}

DEAP-3600 is a single phased liquid argon (LAr) direct dark matter detection experiment that has been collecting data since August 2016 in SNOLAB, in Sudbury, Ontario, Canada. DEAP operates a 3200 kg LAr target mass viewed by 255 photo-multiplier tubes (PMTs) that collect argon scintillation light which has been shifted into the detectable range by a 3$\mu m $ coating of wavelength shifting material 1,1,4,4-Tetraphenyl-1,3-butadiene (TPB) \cite{a}. Electronic recoils (ERs) are differentiated from nuclear recoils (NRs) through pulse shape discrimination (PSD). Nuclear recoils in argon preferentially produce singlet excited states of argon which decay with a lifetime $\tau_{s} \sim 6ns$, while  $\gamma$ and $\beta$ induced ER background events excite argon nuclei preferentially into triplet states that decay with a long lifetime of $\tau_{t} \sim 1500ns$ . By comparing the difference in arrival time of scintillation photons ER (background-like) events can be distinguished from NR (signal-like) events \cite{b}. 

\par
In 2017, collaboration members from the DEAP collaboration, along with members from mini-CLEAN, ArDM and Darkside, formed the Global Argon Dark Matter Collaboration (GADMC) \cite{c}. The collaboration is working on a phased approach with the ultimate goal being the construction of a multi-hundred tonne LAr dark matter detector `Argo' planned to operate at SNOLAB. The next generation detector Argo will have a design sensitivity down to the argon `neutrino floor', the point at which coherent elastic neutrino scattering from argon becomes a significant background. Research and development efforts are underway to this end. 

\par

The Carleton Noble Liquids Detector Lab (COLD Lab) is a detector development research lab and class 10,000 clean room located at Carleton University, Ottawa, Ontario, Canada. Some of the objectives of this facility include the development of detection techniques for noble liquid detectors, development of techniques for cryogenic silicon photo-multiplier (SiPM) readout, as well as studies of scattering and absorption of bulk noble liquids. 
\newpage
\section{Argon - 1 detector}
\label{sec:A1}

As part of the COLD Lab research program a 30 kg LAr test detector has been constructed. Argon - 1 is a small scale single phase LAr detector which will be outfitted with SiPMs for scintillation light detection. The goals of Argon-1 include; characterization of SiPM devices and readout for noble liquid detectors, studies of wavelength shifting materials, and the demonstration of a novel surface background rejection technique using a layered slow scintillating surface material \cite{d}. 

\begin{figure}[h]
    \centering
    \includegraphics[width=0.5\textwidth]{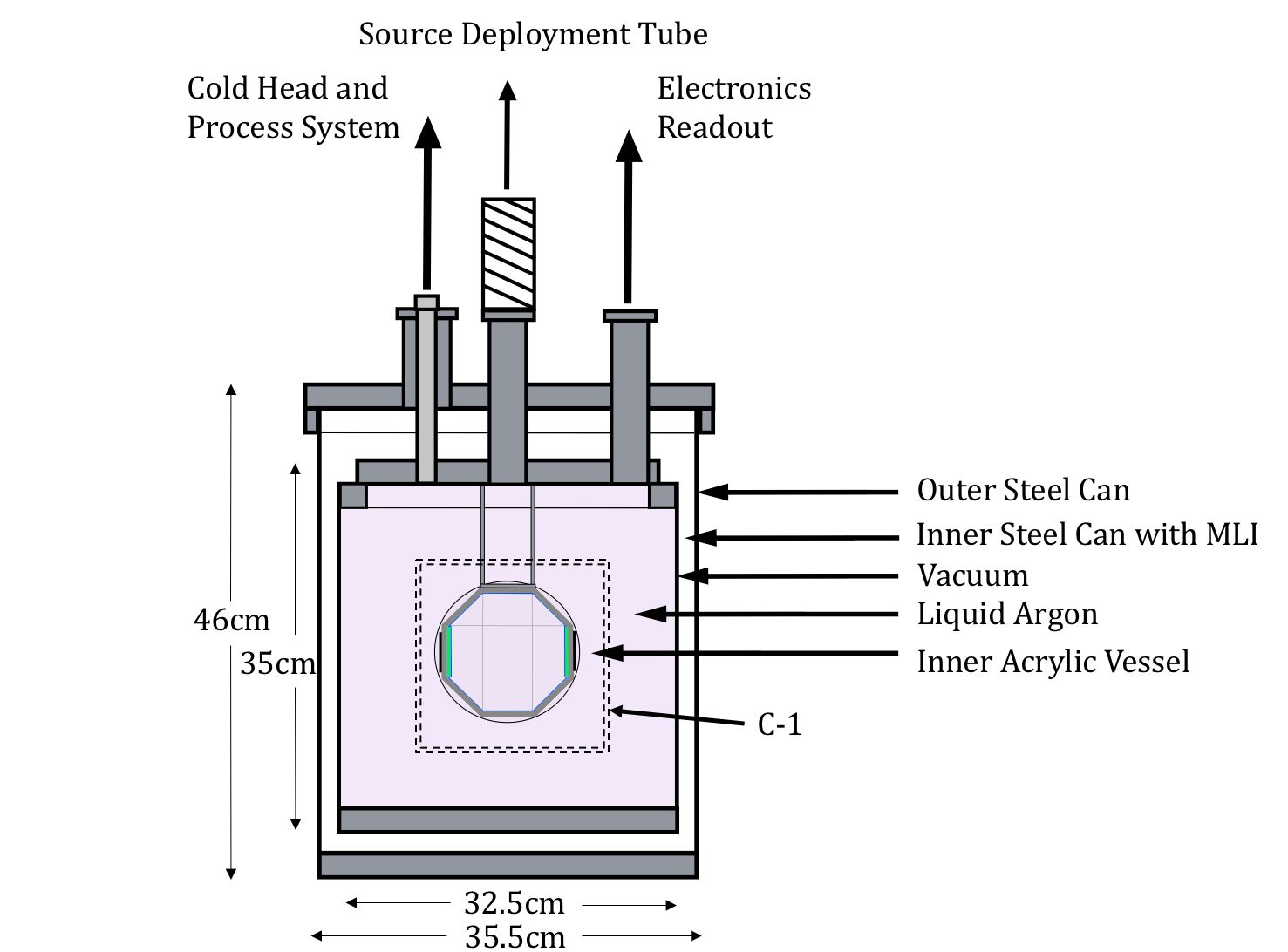}
    \includegraphics[width=.45\textwidth,origin=c]{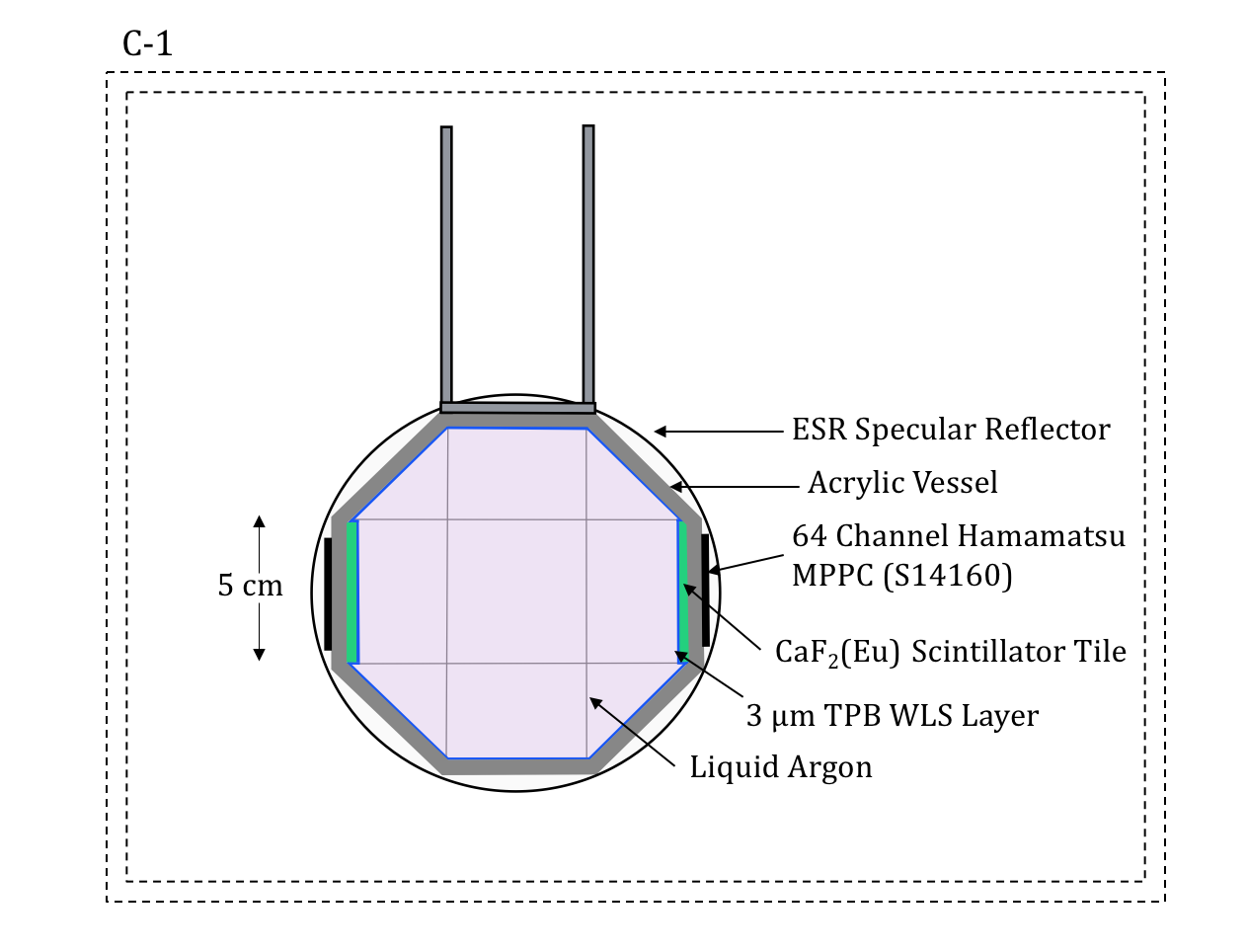}
    \caption{Left: Labeled sketch of Argon-1 detector, Inner vessel is shown as outlined for the first iteration, Right: Section C-1, Inner vessel with two 64 Channel SiPM Arrays and 2 $CaF_{2}$ slow scintillator disks}
    \label{fig:a1_draw}
\end{figure}

\par
The Argon-1 vessel is a 26 sided polyhedron, designed to approximate a spherical vessel for symmetry while being re-configurable for testing of different WLS materials, the flat faced sides also allow for full contact of SiPM windows to the acrylic of the vessel. Both panels with SiPMs will be instrumented with various slow scintillators for testing, starting with $CaF_{2}$ disks. To begin with, the surface of the entire inner vessel will be vacuum evaporated coated with TPB which will serve as the wavelength shifter. The vessel will be surrounded by 3M ESR specular reflective foil and submerged in liquid argon, the inner can is surrounded by multi-layer insulation and vacuum nested inside of an outer stainless steel can. A sketch of the detector is shown in figure (\ref{fig:a1_draw}), the inner vessel is connected to a purification loop which as designed, will take in high-purity argon gas that will be fed through a SAES getter purifier to generate detector-grade purity argon gas, through a radon trap, and then condensed into liquid inside of a copper mounted cooling coil connected to a coldhead. Argon boil off from the inner can is passed back through the process loop for steady state operation.  

\par

The small-scale nature of the detector allows for convenient reconfiguration, and several iterations of the detector are currently planned. The first iteration is described in detail in this text, and future iterations will employ different WLS materials, scintillating materials and higher photodetector coverage. In the first iteration the inner vessel will be viewed by two 64 channel SiPM arrays (Hamamatsu S14161-3050HS-08), with peak sensitivity at 450 nm. Each channel will be read out individually, passed through a pre-amplification stage then into two 64 channel CAEN V1740 digitizers for readout. The high granularity of the SiPM readout will allow for efficient discrimination of surface events from bulk argon events. At the time of writing the detector is operational and will be outfitted with SiPMs and front-end electronics in the coming months.

\section{Surface background rejection technique}
\label{sec:SBRT}
Emanation of radon into sensitive detector materials for low-background experiments constitutes a significant source of backgrounds in the form of $\alpha$ decaying isotopes such as Polonium-210 due to long lived intermediate progeny. Degraded surface $\alpha$'s and recoiling Lead-206 nuclei can mimic low energy nuclear recoils in liquid argon detectors. A common method for the removal of these events is through position reconstruction, where by validation with simulated data the event vertex is reconstructed and any events near the surface of the vessel are removed, or `fiducialized'.The difficulty in modelling complicated surface event topologies results in large fiducialization and reduction of signal acceptance. Surface events may also be discriminated against due to the long lifetime of TPB scintillation from alpha particles through PSD \cite{e}. The proposed technique for surface event rejection through PSD allows for removal of a large background without simulated model-dependent position reconstruction and at a higher efficiency than solely through TPB. 

\par
\begin{figure}[h]
    \centering
    \includegraphics[width=.4\textwidth]{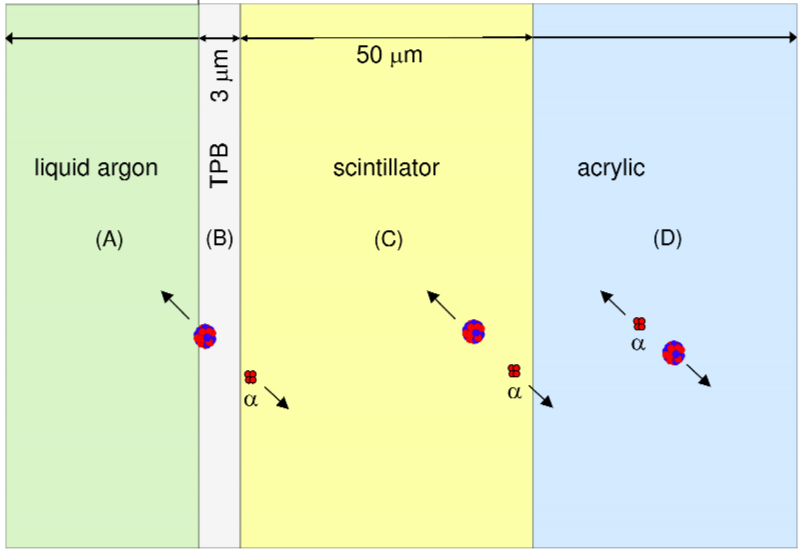}
    \qquad
\includegraphics[width=.4\textwidth,origin=c]{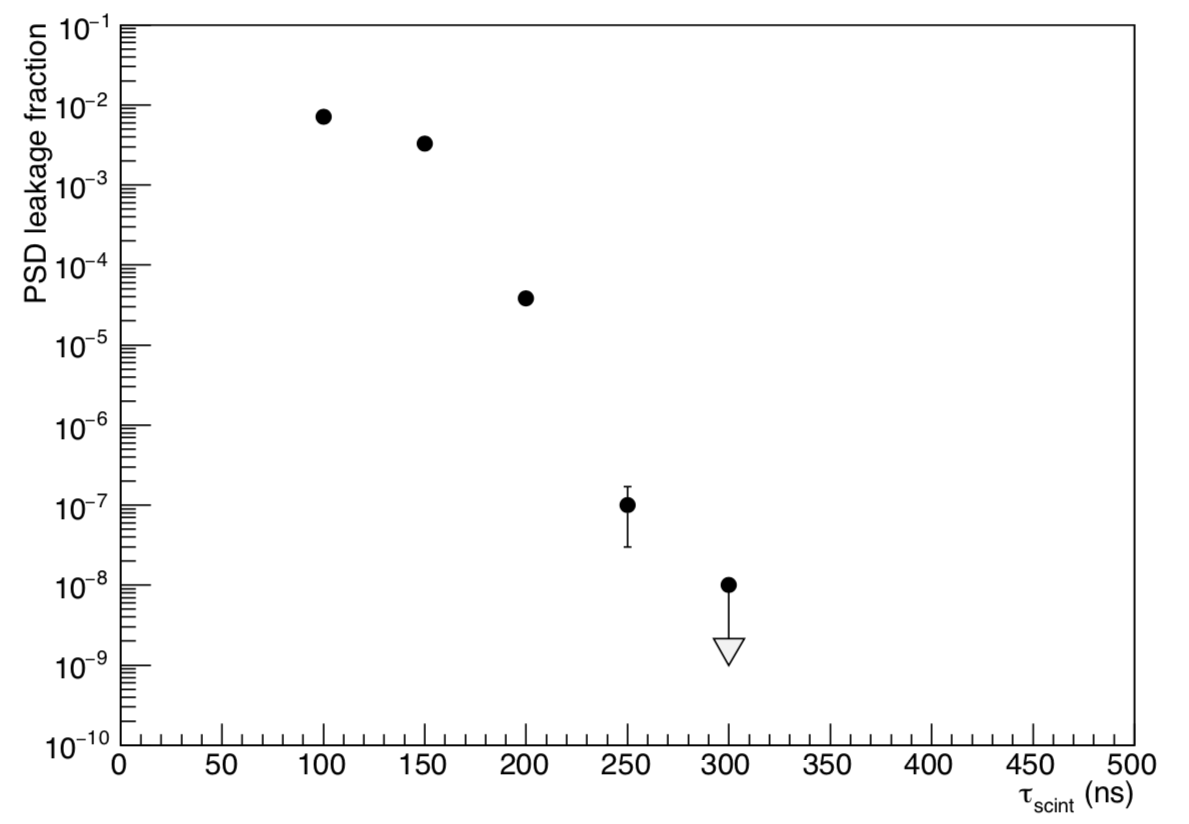}
    \caption{Left: Surface event rejection using a long lifetime scintillating layer , Right: PSD Leakage into a WIMP region of interest in LAr from surface $\alpha$'s as a function of scintillator lifetime from \cite{d}}
    \label{fig:surf}
\end{figure}

By implementing a thin layer of scintillating material a long time constant between the non-active vessel (Commonly acrylic) and wavelength shifting layer (Commonly TPB) as shown in figure (\ref{fig:surf}) the most common surface events can be discriminated against. 
\par
As illustrated in figure (\ref{fig:surf}), there are several common decay modes that can be identified and discriminated against using this technique.
\begin{itemize}
    \item Case 1: $\alpha$ decay events in the WLS layer deposit all of their energy in an active region are identified readily. The high energy of the decay also places them outside of the region of interest.
    \item Case 2: Decays in region C) deposit all or most of their energy in the slow scintillating layer and are identified through PSD, since NR events in LAr are primarily prompt. Selecting a material with a sufficiently long lifetime shifts these events away from the region of interest with efficiency at the $10^{8}$ level as shown in figure (\ref{fig:surf}) right.
    \item Case 3: Decays in non-active region D) that deposit energy in region C) and are likewise shifted from the prompt region of interest or produce no detectable signal.
\end{itemize}  

\par 

The first candidate slow scintillator for Argon-1, $CaF_{2}$, has a lifetime of $\sim 10\mu s$ at liquid argon temperatures \cite{f}, and the relatively high light yield of 4000 photons per MeV of  $\alpha$ energy, and an emission spectrum peaked at 435 nm makes it a suitable candidate material for validation of the proposed technique in a liquid argon environment. Table (\ref{tab:i}) summarizes the relevant scintillation parameters for the $CaF_{2}$, LAr and TPB, light yields for LAr and $CaF_{2}$ are shown with quenching applied for a more direct comparison. 

\begin{table}[htbp]
\centering
\caption{\label{tab:i} Scintillation parameters of active media}
\smallskip
\begin{tabular}{|l|c|c|c|}
\hline
Media & Time constants & Light yield & Peak emission (nm)\\
\hline
$CaF_{2}$ & $\sim\, 10\mu s$  \cite{f} & $4\,ph/keV_{\alpha}$ \cite{g} & 435 \cite{g}\\
TPB & $\sim \, 11\,ns$ \& $\sim 275\,ns$ \cite{h} & $0.8\, ph/keV_{\alpha}$ \cite{h} & 420 \cite{h}\\
LAr & $\sim \, 6\,ns$ \& $\sim \,1500 \,ns$ \cite{b}& $30\,ph/keV_{\alpha}$ \cite{i} & 128 \cite{b}\\
\hline
\end{tabular}
\end{table}
\par
\begin{figure}[h]
    \centering
    \includegraphics[width=0.65\textwidth]{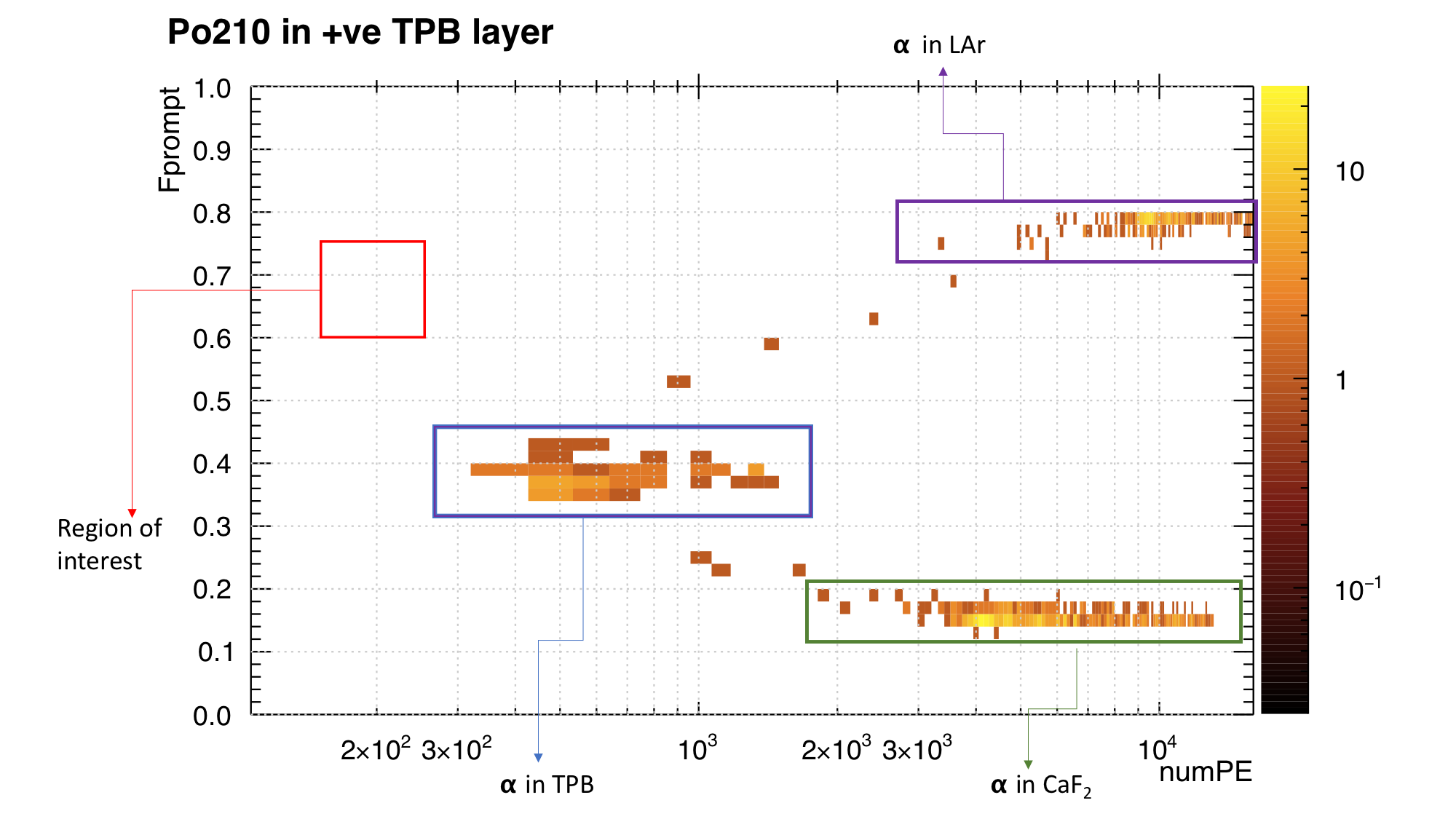}
    \caption{Simulation results showing `Fprompt' PSD estimator versus `numPE', an energy estimator using number of detected photo-electrons, Po-210 $\alpha$ decay events are simulated uniformly in a 3$\mu$m TPB layer, various populations of events are outlined as well as the approximate low energy Ar-40 NR(WIMP-like) ROI}
    \label{fig:po210}
\end{figure}
\begin{figure}[h!]
    \centering
    \includegraphics[width=.4\textwidth]{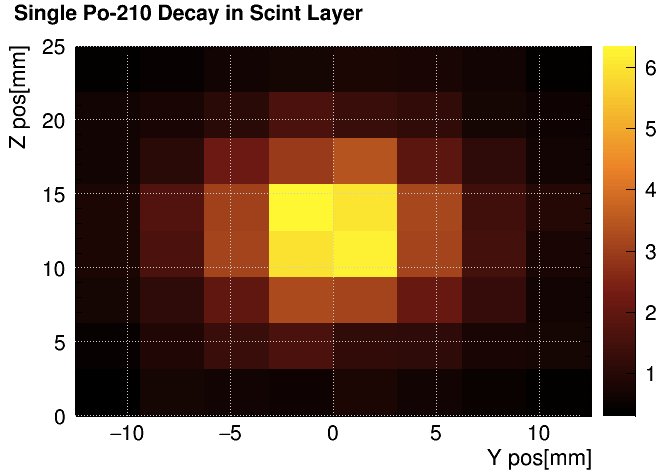}
    \qquad
\includegraphics[width=.44\textwidth,origin=c]{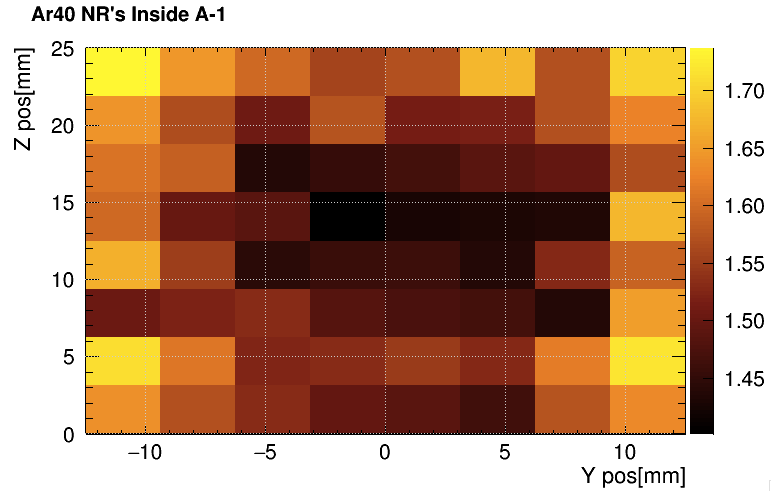}
    \caption{Left: Simulated SiPM response to Po-210 Decays in the scintillator layer, each bin corresponds to a single SiPM channel with percent of total light in an event plotted on the z-axis, Right: Response from Ar-40 NR events in the LAr bulk}
    \label{fig:sipm_response}
\end{figure}

A GEANT4 based simulation of this technique using the software framework RAT is in development, and preliminary simulation results are shown in figures (\ref{fig:po210}) and (\ref{fig:sipm_response}).

`Fprompt' shown in eq. \ref{fprompt} is the PSD variable used, and is defined as the fraction of total light in a prompt window [-50ns,60ns]. Figure (\ref{fig:po210}) shows Fprompt  versus the energy of an event in `numPE'; the Monte Carlo truth value of detected photo-electrons, for Po-210 $\alpha$ decay events in the TPB WLS layer.

\begin{equation}
\label{fprompt}
Fprompt = \frac{\sum_{t=-50ns}^{60ns}PE(t)}{\sum_{t=-50ns}^{10\mu s}PE(t)}
\end{equation}

By implementing a thin $CaF_{2}$ layer between TPB and acrylic layers and simulating Po-210 events in the TPB layer the energy of the potential region of interest events is shifted upwards, and the events are also shifted down in Fprompt with large separation from the region of interest. This is depicted in figure (\ref{fig:surf}) by case 1.

\par

\par
High granularity SiPM arrays allow for the discrimination of bulk argon events from surface events, shown in figure (\ref{fig:sipm_response}). Left is the response from a Po-210 decay in the scintillating layer, and right shows the response from Ar-40 NR events from the inner bulk of the detector. Surface events have categorically higher concentrations of light in the SiPM at the interface due to solid angle effects. The quantification of light concentration can be estimated through eq. \ref{fmaxpe}.

\begin{equation}
\label{fmaxpe}
   fLight = \frac{PE \:in \:brightest\:channel}{Total \:PE} 
\end{equation}
Utilization of this difference will allow for tagging of surface events in data to demonstrate the efficiency of the proposed technique with the Argon-1 detector.

\section{Outlook}

Future low-background liquid argon detectors can benefit greatly from a strong surface background rejection technique without reliance on position reconstruction. The proposed technique is projected to allow discrimination of surface events with an efficiency of $> 10^{8}$ for a scintillating layer with a lifetime greater than 300$ns$. Argon-1, a test detector located at Carleton University, has been constructed and is being outfitted with high granularity SiPM arrays which will provide the capability to test and validate the proposed novel background surface background rejection technique described here. Determination of the technique's rejection efficiency as well as the study of candidate WLS materials are among the goals of Argon-1, and progress has been made working towards commission of a fully operational detector with 128 Channel SiPM readout along with a full scale GEANT4 based simulation for verification of results.

\acknowledgments
This research has been supported by the Canada Foundation for Innovation, the Natural Sciences and Engineering Research Council of Canada, the Arthur B. McDonald Canadian astroparticle physics research institute and the Ontario Research Fund.


\end{document}